\documentclass[twocolumn,showpacs,preprintnumbers,amsmath,amssymb,prl]{revtex4}%
\usepackage{graphicx}
\usepackage{dcolumn}
\usepackage{bm}
\usepackage{times}
\usepackage{amsmath}
\usepackage{amsfonts}
\usepackage{amssymb}%
\providecommand{\U}[1]{\protect\rule{.1in}{.1in}}

\begin{document}
\preprint{S. J. Moon \textit{et al.}}
\title{Dual Character of Magnetism in Ferropnictides: Insights from Optical Measurements}
\author{S.J. Moon,$^{1}$ J.H. Shin,$^{1}$ D. Parker,$^{2,*}$ W.S. Choi,$^{1}$ I.I. Mazin,$^{2}$
Y.S. Lee,$^{3}$ J.Y. Kim,$^{4}$ N.H. Sung,$^{4}$ B.K. Cho,$^{4}$
S.H. Khim,$^{5}$ J.S. Kim,$^{6}$ K.H. Kim,$^{5}$ and T.W
Noh$^{1,\dag}$} \affiliation{$^{1}$ReCOE $\&$FPRD,\;Department of
Physics and Astronomy, Seoul National University, Seoul 151-747,
Korea} \affiliation{$^{2}$Naval Research Laboratory, 4555 Overlook
Avenue, SW, Washington, D.C. 20375, USA}
\affiliation{$^{3}$Department of Physics, Soongsil University,
Seoul 156-743, Korea} \affiliation{$^{4}$Department of Nanobio
Materials and Electronics $\&$ Department of Materials Science and
Engineering, GIST, Gwangju 50-712, Korea} \affiliation{$^{5}$CSCMR
$\&$FPRD,\;Department of Physics and Astronomy, Seoul National
University, Seoul 151-747, Korea} \affiliation{$^{6}$Department of
Physics, POSTECH, Pohang 790-784, Korea }
\date{\today}

\begin{abstract}
We investigate the electronic structure of EuFe$_{2}$As$_{2}$
using optical spectroscopy and first-principles calculations. At
low temperature we observe the evolution of \textit{two} gap-like
features, one having a BCS mean-field behavior and another with
strongly non-BCS behavior. Using band structure calculations, we
identify the former with a spin-Peierls-like partial gap in
$d_{yz}$ bands, and the latter with the transition across the large
exchange gap in $d_{xz}/d_{xy}$ bands. Our results demonstrate that the
antiferromagnetism in the ferropnictides is neither fully local nor
fully itinerant, but contains elements of both.
\end{abstract}

 \maketitle

The discovery of high-temperature superconductivity in the
ferropnictides has stimulated great interest in investigating
their physical properties \cite{Kamihara}. As in the cuprates, the
superconducting state of the ferropnictides is located in the
vicinity of a magnetically ordered state. The parent ferropnictide
compounds exhibit stripe-type antiferromagnetic (AFM)
spin-density-wave (SDW) order \cite{Cruz, RotterPRB, Zhao1,
Huang}, while suppressing the AFM order by various means gives
rise to the superconductivity \cite{Kamihara, Chen, RotterPRL,
Torikachvili, Takahashi}. It is generally believed that the
proximity to magnetism plays a crucial role in establishing
electronic pairing \cite{MS}.

Numerous efforts have been made to elucidate the electronic
structure changes across the SDW transition of the ternary
ferropnictide $A$Fe$_{2}$As$_{2}$ ($A$=Ba, Sr, or Eu) compounds.
In angle-resolved photoemission spectroscopy (ARPES) studies on
BaFe$_{2}$As$_{2}$ and SrFe$_{2}$As$_{2}$, the electronic
structure changes across the SDW transition were initially
explained in terms of the exchange splitting \cite{Yang, Zhang},
even though no uniform exchange splitting is expected in an
antiferromagnet. Later ARPES studies, optical and quantum
oscillation measurements as well as transport data indicated
substantial, albeit incomplete, gapping of the Fermi surfaces due
to the SDW gap opening \cite{Hsieh, Hu, Wu, Pfuner, Sebastian,
Wen}. In particular, an optical study by Hu \textit{et al}. showed
the appearance of two gap-like features (GLFs) in the SDW states of BaFe$_{2}$As$_{2}$ and SrFe$_{2}$As$_{2}%
$, reflecting the multiband nature of the ferropnictides
\cite{Hu}. However, Wu \textit{et al}. reported that the optical
conductivity spectra $\sigma(\omega)$ of EuFe$_{2}$As$_{2}$ showed
only one GLF in the SDW state \cite{Wu}. Hsieh \textit{et al}.,
using polarization-dependent ARPES on SrFe$_{2}$As$_{2}$, reported
two sets of Fermi surfaces, nested and non-nested, and concluded
that only the nested bands exhibited a partial gap opening in the
SDW state \cite{Hsieh}.

The small crystal field splitting and large bandwidth of the
ferropnictide compounds make multiband effects crucial in the
formation of the electronic structure both in the vicinity of the
Fermi energy and at higher energies \cite{Singh,Ma,Mazin}. This
complicates the interpretation of the electronic structure.
Moreover, particularly in the 122 family, the large magnetic
moments cause reconstruction of the electronic bands in the SDW
state not only in the vicinity of the Fermi level (as a simplistic
nesting picture would imply), but over the entire bandwidth. The
multiband and multi-orbital nature of the electronic structure
suggests an interesting interplay between the orthorhombic
anisotropy of the magnetic order (even neglecting the small
structural distortion) and the orbital character of the bands near
the Fermi level. However, to date, few studies have addressed the
effect of magnetism on the multiple electronic bands. In light of
this, understanding the changes of the individual electronic bands
due to magnetism, and the effects of these changes on
spectroscopic quantities such as optical conductivity, is of great
interest.

In this Letter, we investigate the evolution of the electronic
structure of EuFe$_{2}$As$_{2}$ using optical spectroscopy and
first-principles density-functional calculations. We observe a
drastic decrease of the Drude response and the formation of two
GLFs in the $\sigma(\omega)$ across the SDW transition. The GLFs
exhibit disparate temperature ($T$) dependences, with one GLF
magnitude and spectral weight (SW) in quantitative agreement with
the BCS weak-coupling behavior, while the other is much less
$T$-dependent. Based on first-principles calculations, we identify
the two corresponding classes of optical transitions: one is
related to the transition within the spin-minority bands and is
therefore sensitive to the long-range magnetic order, and the
other between the spin-majority and the spin-minority (on the site
with the opposite magnetization) states. The second transition is
mostly defined by the local magnetic moment, and less sensitive to
the long-range ordering.
\begin{figure}[ptb]
\includegraphics[width=3in]{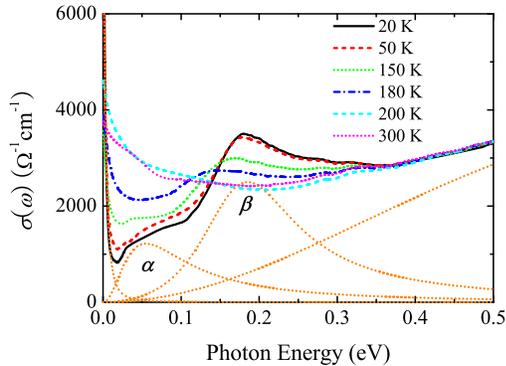}
\vspace{-0.4cm}
\caption{(color online). In-plane
$\sigma(\omega)$ of EuFe$_{2}$As$_{2}$. The orange dotted lines represent the
contributions of the Drude, peak $\alpha$, peak $\beta$, and a high energy
interband optical transition in the Drude-Lorentz oscillator fit at 20 K.}%
\end{figure}

High-quality EuFe$_{2}$As$_{2}$ single crystals were grown using the flux
technique. The resistivity of EuFe$_{2}$As$_{2}$ showed an anomaly at $T_{N}%
$=190 K, which is associated with the paramagnetic (PM)-to-AFM
transition. Near-normal incident reflectance spectra $R$($\omega$)
were measured between 5 meV and 6.5 eV, using a liquid-He cooled
cryostat. The in-plane $\sigma(\omega)$ were obtained using the
Kramers-Kronig transformation of $R$($\omega$). In the
calculations, the experimental tetragonal and orthorhombic lattice
constants were used for the PM and AFM states, respectively
\cite{Tegel}.

Figure 1 displays the $T$-dependent $\sigma(\omega)$ of EuFe$_{2}%
$As$_{2}$. As $T$ decreases across the SDW transition, the broad
Drude-like response becomes strongly suppressed and develops into
a gap-like structure. These $T$-dependent $\sigma(\omega)$ of
EuFe$_{2}$As$_{2}$ reveal a couple of intriguing features. First,
in the SDW state, there is a very sharp Drude response in the
frequency region below about 0.02 eV. Its small SW and even
smaller width, are consistent with a large reduction of the free
carrier concentration, due to the opening of a partial gap at the
Fermi surface, and an even stronger reduction of the scattering
rate, as noticed before \cite{Wang, Fang, JSLee}. Second, unlike
the weight transfer for $\sigma(\omega)$ of single-band density
wave systems, two GLFs (peaks $\alpha$ and $\beta$) appear below
the SDW transition that receive the SW lost due to the gapping of
the Fermi surface. This clearly reflects the multiband nature of
the ferropnictides.

The evolution of these two peaks with $T$ provides insight into
the characteristic responses of the individual bands to magnetism.
To obtain such information, we analyzed the $\sigma(\omega)$ using
the Drude-Lorentz oscillator model. Overall, we used two Drude
($D$) terms and three Lorentz oscillators for peak $\alpha$, peak
$\beta$, and a high energy interband transition. The second Drude
peak is not necessary for fitting below $T_{N},$ but if we include
it, it comes out with a smaller weight and a very large width,
and, consequently, with a large uncertainty. According to the Hall
data \cite{Wen}, the hole mobility in the SDW state is drastically
smaller than the electron mobility. Thus, we tentatively attribute
the first, narrow Drude peak to electrons, and the second, broad
and with the SW corresponding to 1.5-2 times smaller Fermi
velocity, to holes. Note also that above $T_{N}$ the intensity of
the low-energy peak $\alpha$ is zero within the fitting
uncertainty.

Next we turn to the peak $\alpha$, which is well-defined below
$T_{N}.$ We find $2\Delta_{\alpha}/k_{B}T_{N}\sim$3.4, where
$\Delta_{\alpha}$ is taken equal to the peak position
$\omega_{\alpha}$ at 20 K. Note that this value agrees well with
the mean-field BCS ratio. Also, the $T$ dependences of
$\omega_{\alpha}$ and SW$_{\alpha}$ follow the BCS formula quite
well, as shown with dashed lines in Figs. 2(a) and 2(b).
Conversely, for the peak $\beta$, the 2$\omega_{\beta}$(20
K)$/k_{B}T_{N}$ value is about 10.9, and there are hardly any $T$
dependence in $\omega_{\beta}$ and SW$_{\beta}$. Above $T_{N}$,
the SW$_{\beta}$ abruptly decreases, but its position is about the
same as that below $T_{N}$. Importantly, the total SW is
reasonably well conserved across the transition, with the SW
removed from the Drude peaks in the SDW phase, and transferred
into the GLF $\alpha$ and $\beta$. The overall picture suggests
that the peak $\alpha$ is associated with partial gapping that
evolves in a mean-field manner, while the bands related to the
peak $\beta$ are almost fully gapped and the evolution is a more
local phenomenon and involves wholesale alteration of the band
structure.

\begin{figure}[ptb]
\includegraphics[width=8cm]{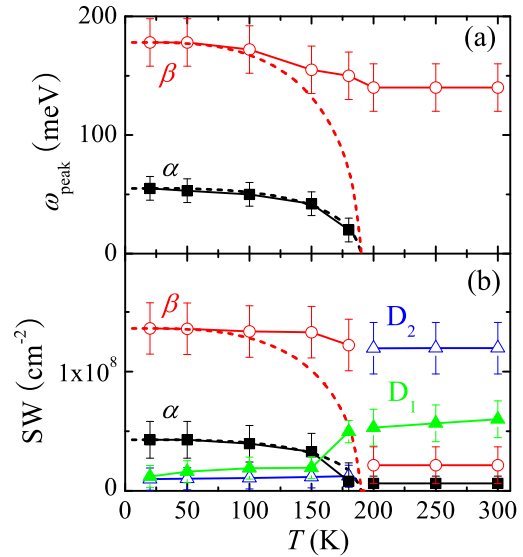}
\caption{(color online). $T$-dependent changes in (a) the energy
values of the peaks $\alpha$ and $\beta$ and (b) the SW of the
peaks $\alpha$ and $\beta$, as well as the SW of the two Drude
peaks. The dashed lines show the $T$-dependences of the peak
energies and SW according to the BCS theory for the peaks $\alpha$ and $\beta$.}%
\end{figure}

This can be compared with Hsieh \textit{et al}.'s suggestion that two
different kinds of bands, nested and non-nested bands, exist in SrFe$_{2}%
$As$_{2}$ \cite{Hsieh}. They argued that the nested bands
experience partial gapping, but that the non-nested bands were
assumed to have no substantial change upon entering the SDW state.
However, the SDW gap these authors observed in their ARPES
experiment was rather small and not consistent with results from
our (and others) optical studies (and also too small for the
observed SDW magnitude of $\sim1$ $\mu_{B})$. As shown in Fig.
2(b), most of the carriers present above $T_{N}$ at the Fermi
level disappear below $T_{N}$, and 90\% of the Drude SW is shifted
into the interband peaks $\alpha$ and $\beta$. The discrepancy
with the ARPES results most likely is caused by the sensitivity of
ARPES (but not optics) to the surface states, which likely have a
smaller magnetic moment than in the bulk.

To directly identify the electronic structure changes in EuFe$_{2}$As$_{2}$
due to the SDW transition, we performed all-electron density-functional
calculations using the local density approximation, as implemented in the
WIEN2K code. To simulate the actual small experimental magnetic moment of
$\sim1\mu_{B}$, we applied an artificial negative $U$ of $-$0.03 Ry, which
reduced the calculated staggered magnetization from $1.65\mu_{B}$ to
$1.15\mu_{B}$ (this procedure has no physical justification and is only
employed to emulate the reduction of the ordered moment due to quantum fluctuations).
\begin{figure}[h!]
\includegraphics[width=8cm]{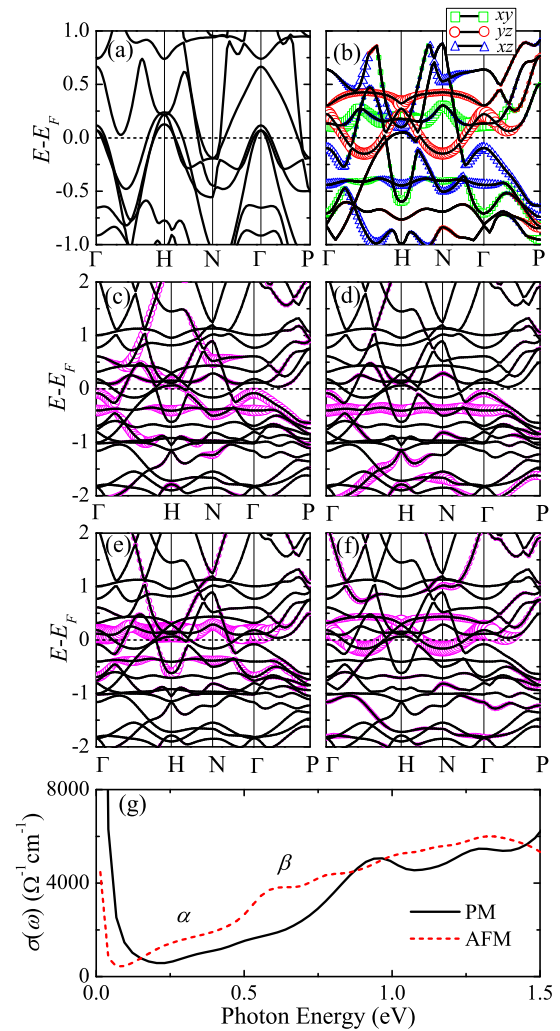}
\caption{(color online). (a) Band structure of the PM tetragonal
state. (b) Band structures of the AFM orthorhombic state. The
contributions of the $xy$, $yz$, and $xz$ orbital states are
depicted by the open squares, circles, and triangles. The size of
the symbol indicates the fraction of the orbital character of
bands. The symmetry points are $\Gamma$=(0, 0, 0), H=(2$\pi$, 0,
0), N=($\pi$, $\pi$, 0), and P=($\pi$/2, $\pi$/2, $\pi$/2) in the
PM Brillouin zone for (a) and AFM Brillouin zone for (b)-(f). Note
that the AFM Brillouin zone is rotated by $\pi$/4 relative to that
of the tetragonal state and is downfolded. (c) The contribution of
Fe-1 $d_{xz}$ spin-up character. (d) The contribution of Fe-1
$d_{xz}$ spin-down character. (e) The Fe-2 $d_{xy}$ spin-down
character. Here the Fe-2 atom is adjacent to Fe-1 in the direction
of the ordering vector. (f) The Fe-1 $d_{yz}$ spin-up character.
(g)
Calculated in-plane $\sigma(\omega)$ for the PM and AFM states.}%
\end{figure}

Figure 3(a) shows the band structure of the PM state in the energy
region between $-1$ and 1 eV, where the bands have predominantly
Fe $d$ character. Near the zone center $\Gamma=(0,0,0)$, three
hole bands cross the Fermi level $E_{F}$, while near the
$N=(\pi,\pi,0)$ point, two electron bands cross the $E_{F}$. As
shown in Fig. 3(b), the band structure changes drastically upon
entering the AFM state. In the AFM state, there are far fewer
$E_{F}$ crossings, and as is evident from Fig. 3(b), these are
almost entirely of $d_{yz}$ character, which is consistent with
recent results from a laser-ARPES study \cite{Shimojima}. This
therefore reflects the gapping of the bands of $d_{xy}$ and
$d_{xz}$ character. Here $x$ is the direction of the longer (FM)
Fe-Fe bond and $y$ the direction of the shorter (AFM) bond.

Evidently, magnetism affects the band structure on the energy
scale of several eV. A comparison of the Fe $d_{xz}$ spin-up and
spin-down band character plots in Figs. 3(c) and 3(d) reveals that
the spin-up bands of this character lie approximately 1 eV higher
than the spin-down bands, and this energy difference carries over
to the bands with other orbital characters (not shown) as well.
This energy difference can be quantitatively interpreted in terms
of exchange splitting, based on the local Fe Hund coupling of
$\sim$0.8 eV and the calculated staggered magnetization of
$1.15\mu_{B}$. This exchange splitting, not the Hubbard $U$ or
Fermi surface nesting, drives the main band-structure changes
across the ordering transition.

Moreover, these band-structure changes and Hund coupling can be
directly related to the features at about 0.2 eV and 0.055 eV in
the measured $\sigma(\omega)$. Presented in Figs. 3(d) and 3(e)
are bands with the Fe $d_{xz}$ spin-down character projection and
the $d_{xy}$ projection for the same spin on an
antiferromagnetically aligned adjacent Fe, respectively. The
latter is identical by symmetry to the spin-$up$ projection on the
$same$ Fe, therefore they are separated by roughly the exchange
splitting.

We immediately notice a flat band in the former projection at
approximately $-$0.4 eV and the latter at 0.2 eV, suggesting the
existence of an 0.6 eV feature, which is clearly observed in the
calculated $\sigma(\omega)$ (Fig. 3(g)). This feature results from
transitions between the majority occupied Fe spin-down $d_{xz}$
orbital \textit{on one Fe atom}, Fe-1, and the minority unoccupied
spin-down $d_{xy}$ orbital \textit{on an adjacent Fe atom} with
the opposite spin, Fe-2. The parallel nature of the bands then
results in a large contribution to the joint density of states
(JDOS) at 0.6 eV, which drives the larger gap. (There is of course
a similar effect linking the occupied spin-up Fe-2 orbital state
with the unoccupied spin-up Fe-1 orbital state.) As this
interaction occurs in real space, it is sensitive to the local
exchange splitting, and is therefore the harbinger of localized
moments. Factoring in the band mass renormalization in the
pnictides of approximately 3, as measured through ARPES \cite{Lu},
one concludes that this is the peak $\beta$ at about 0.2 eV in the
measured $\sigma(\omega)$. Supporting this assignment is the fact
that this peak is experimentally observed to have little $T$
dependence in the SDW state and to change abruptly its weight
(number of instant AFM bonds), but not its position (local moment
magnitude) across the magnetic transition.

The origin of the peak $\alpha$ at about 0.055 eV in the measured
$\sigma(\omega)$ is somewhat more complex. The $d_{yz}$ spin-down
band in Fig. 3(f) is the one that provided metallic carriers at
the $E_{F}$. In the plot of the $d_{xz}$ spin-up character (Fig.
3(c)) one sees a band parallel to this band, but displaced down by
approximately 0.25 eV. These parallel bands provide a distinct
contribution to JDOS and result in a feature at 0.25 eV in the
calculated $\sigma(\omega)$. Unlike the larger gap, however, this
gap originates from transitions between two same-spin orbitals
within\textit{ the same-spin }sublattice, and is therefore
insensitive to the local exchange splitting (we have verified that
reducing the local moment in the calculations suppresses the
frequency of the higher-energy peak, but affects little the
lower-energy feature). This gap is thus an itinerant-type SDW one,
despite substantial band-structure effects. Considering the mass
renormalization factor, we conclude that this smaller peak is the
peak $\alpha$ seen at 0.055 eV in the experiment, which obeys
markedly mean-field BCS-type behavior.

The magnetism in the ferropnictides is thus neither fully local
nor fully itinerant, but has elements of both, as discussed in
Ref. \onlinecite{JM}. Our experiment shows that this duality
translates into two distinctively different sets of optical
features. One is a low-energy feature that monitors the true long
range order manifesting itself in restructuring the Fermi surface,
and thereby gapping out the Drude SW. This spatial ordering is
comparable to the conventional spin-Peierls SDW, except it
develops on the background of the large local moments, which
produce substantial band-structure changes. The higher energy
feature is less sensitive to the long-range spatial order, but
rather to the local exchange splitting. Intermediate-range
ordering is sufficient to slow down the local fluctuations enough
to make the local splitting visible in optics.

In summary, we have investigated the electronic structure of EuFe$_{2}$%
As$_{2}$ using optical spectroscopy and first-principles
calculations. We observe drastic changes of the optical
conductivity across the magnetic transition, which can be
described as transfer of the spectral weight into two gap-like
features. These are very sensitive to the symmetry breaking
associated with the stripe antiferromagnetism. The higher-energy
feature is associated with the optical transitions between two Fe
atoms of the opposite spin. The scale is set by the local exchange
splitting, therefore this transition shows little temperature
dependence below $T_{N}$ and survives (with a reduced spectral
weight) above $T_{N}.$ The low-energy feature is associated with
the transitions within the spin-minority states of the same-spin
Fe's. These are less sensitive to the magnitude of exchange
splitting, but very sensitive to the long range order controlling
the gapping of the Fermi surface.  Thermally this feature behaves
as a weak-coupling SDW order parameter, that is, according to the
BCS theory. We conclude that magnetic ordering in pnictides is a
complex two-level process with some states behaving as local and
others as itinerant, and the models using only one side of this
combination are necessarily incomplete. This applies to both
theoretical constructions and interpretations of spectroscopic
experiments.

\indent We acknowledge valuable discussions with S.Y. Park and
H.J. Choi. This research was supported by Basic Science Research
Program through the National Research Foundation of Korea funded
by the Ministry of Education, Science and Technology (MEST) (No.
2009-0080567). The work at GIST was supported by the Korean
Science and Engineering Foundation grant funded by MEST (No.
2009-0078928). KHK is supported by the NRL program (M10600000238)
by MEST.

$^{*}$davidspa@dave.nrl.navy.mil

$^{\dag}$twnoh@snu.ac.kr


\end{document}